\documentclass[superscriptaddress,amsmath,amssymb,
aps, prstab, twocolumn, groupedaddress]{revtex4-2}
\usepackage{amsmath}
\usepackage{natbib}
\usepackage{float}
\usepackage{amssymb}
\usepackage{dcolumn}
\usepackage{graphicx}
\usepackage{longtable}
\usepackage{bm}
\usepackage{color}
\usepackage{epsfig}
\usepackage{xr-hyper}
\usepackage{hyperref}
\usepackage[final]{microtype}  
\usepackage{siunitx}
\usepackage{enumerate}

\newcommand{\TbCrGe}{TbCr$_6$Ge$_6$}

\begin{document}
\title{Wiedemann–Franz violation and thermal Hall effect in Kagome metal TbCr$_6$Ge$_6$} 
\author{Jhinkyu Choi$^{1}$}
\email{choi580@purdue.edu}
\author{Mohan B. Neupane$^{1}$}
\email{J.C., L.L., and M.N. share co-first authorship}
\author{L. H. Vilela-Leão$^{1,2}$}
\email{J.C., L.L., and M.N. share co-first authorship}
\author{Bishnu P. Belbase$^{1}$}
\author{Arjun Unnikrishnan$^{1,3}$}
\author{Syeda Neha Zaidi$^{1}$} 
\author{Jukka I. V\"ayrynen$^{1}$}
\author{Arnab Banerjee$^{1}$}
\email{arnabb@purdue.edu}
\affiliation{$^1$ Department of Physics and Astronomy, Purdue University, West Lafayette, Indiana 47906, USA}
\affiliation{$^2$ Núcleo Interdisciplinar de Ciências Exatas e da Natureza, Universidade Federal de Pernambuco, 55014-900 Caruaru, PE, Brazil}
\affiliation{$^3$ Solid State and Structural Chemistry Unit (SSCU), Indian Institute of Science, Bengaluru - 560012, India}

\begin{abstract}
\indent
The thermal Hall effect has emerged as a powerful probe of exotic excitations in correlated quantum materials, providing access to charge-neutral heat carriers that remain invisible to electrical transport. To directly examine how heat and charge respond in relation within a kagome metal, we investigate the ferrimagnetic rare-earth 1–6–6 compound TbCr$_6$Ge$_6$ using the Wiedemann–Franz (WF) framework. 
We observe a dramatic breakdown of the WF law across the ferrimagnetic transition, where both longitudinal and transverse Lorenz ratios, $L_{xx,xy}=\kappa_{xx,xy}/(T\sigma_{xx,xy})$, deviate strongly from the Sommerfeld value $L_0$. After a partial recovery toward $L_0$ near 5–7 K, the Lorenz ratios are sharply suppressed well below $L_0$ despite a metallic charge response. We further find a pronounced low-temperature suppression of both $L_{xx}$ and $L_{xy}$ and a sign-changing transverse Lorenz ratio, indicating a clear decoupling between heat and charge transport and signaling substantial contributions from charge-neutral excitations whose Berry-curvature-driven transverse response evolves with temperature and magnetic field. 
TbCr$_6$Ge$_6$ thus provides a tunable metallic platform in which exchange-driven ferrimagnetism governs both longitudinal and transverse thermal responses, enabling controlled departures from Wiedemann–Franz behavior over an experimentally accessible temperature and field range.
\end{abstract}

\maketitle

\indent
Kagome metals have emerged as a fertile platform for exploring correlated and topological quantum phenomena, characterized by flat electronic bands, Dirac-like dispersions, and frustrated magnetism~\cite{yin2022topological,li2025electron,ye2018massive,kang2020dirac,liu2018giant}. Their corner-sharing triangular geometry produces destructive interference and large Berry curvature, leading to unconventional charge and heat transport such as anomalous and thermal Hall effects~\cite{nakatsuji2015large,belbase2023large,zhang2024large, xu2022chargeentropy}. In parallel, the thermal Hall effect has become a sensitive probe of charge-neutral excitations, ranging from Bogoliubov quasiparticles in high-$T_c$ and Fe-based superconductors~\cite{Krishana1995YBCO_mfp,Zhang2001YBCO_enhancedKxy,Checkelsky2012BaKFe2As2} to magnons in frustrated magnets~\cite{Onose2010MagnonHall,Hirschberger2015FrustratedMagnet} and possible Majorana modes in Kitaev quantum spin liquids~\cite{Kasahara2018MajoranaQuantization,Yokoi2021HalfIntegerRuCl3,Bruin2022RobustHalfQuantization, Czajka2021RuCl3QO, Czajka2023PlanarThermalHall,czajka2021oscillations}. More recently, sizeable phonon thermal Hall signals in cuprates and SrTiO$_3$~\cite{Uehara2022PhononThermalHallMetallicSpinIce,Grissonnanche2020ChiralPhonons,Li2020SrTiO3PhononTH} have shown that even lattice excitations can produce strong transverse heat currents. These developments highlight that multiple, competing neutral channels may contribute to $\kappa_{xy}$ and that, in Kagome metals, careful comparison of thermal and electrical transport can provide a particularly strong probe of the dominant heat carriers~\cite{Katsura2010QMagnetTheory,Han2019UndopedCuprates,Samajdar2019SquareNeel,Chen2020NearFerroelectric,Guo2020GaugeThermalHall, Qiang2023WFkagome}.

Among metallic kagome systems, the $R T_6 X_6$ family ($R$ = rare earth, $T$ = transition metal, $X$ = Ge or Sn) provides the opportunity to study the coupling of itinerant $d$ electrons in the transition-metal kagome layers with localized $4f$ moments on the rare-earth sites~\cite{brabers1994magnetic,schobinger1997ferrimagnetism,zhang2022electronic,lee2023interplay}. Recent structural characterisation of the $R$Cr$_6$Ge$_6$ series confirms the hexagonal $P6/mmm$ structure of the HfFe$_6$Ge$_6$-type ~\cite{Romaka2024,yang2024crystal, konyk2020electrical}. 
In these materials, itinerant $3d$ conduction electrons in the transition-metal kagome layers are exchange-coupled to localized $4f$ moments on the rare-earth sublattice, so magnetic order can strongly modify both the electronic bands and low-energy spin excitations~\cite{riberolles2023orbital}. In several kagome magnets, such coupling has been shown to generate low-energy magnon modes or spin textures and to produce large anomalous and thermal Hall responses~\cite{higo2018large, zhang2024large,onose2010observation,Hirschberger2015,Laurell2018,Zhuo2021}.
In the $R$Cr$_6$Ge$_6$ subclass with a Cr kagome network, band-structure calculations and spectroscopy reveal partially flat bands near the Fermi level in YCr$_6$Ge$_6$~\cite{ishii2013ycr6ge6,yang2024crystal}.

\begin{figure*}[t]
  \includegraphics[width=1\textwidth]{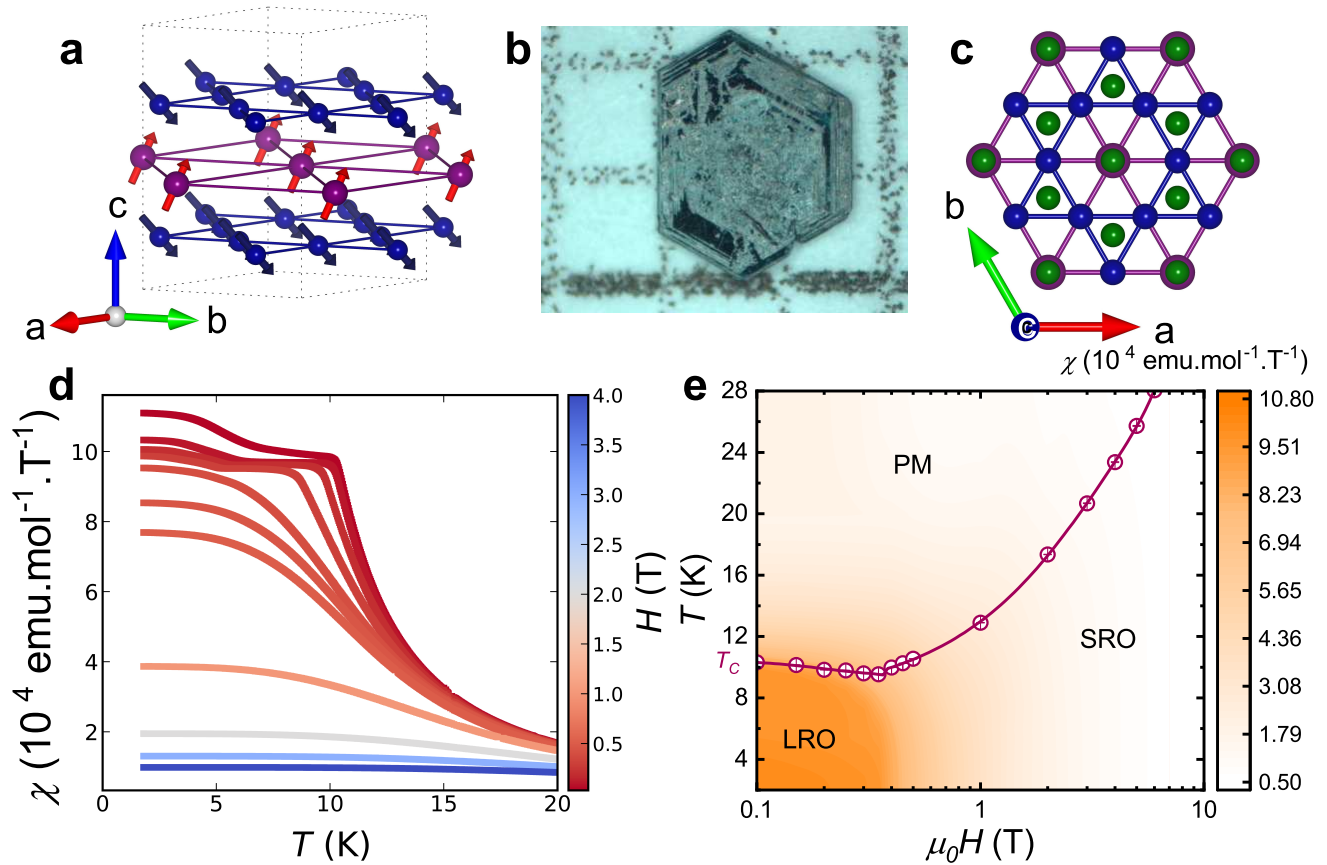}
  \caption{\textbf{Crystal structure and phase diagram of \texorpdfstring{TbCr$_6$Ge$_6$}{TbCr6Ge6}.}
    \textbf{a.} Crystal and magnetic structure of TbCr$_6$Ge$_6$ (HfFe$_6$Ge$_6$-type, $P6/mmm$) showing Cr kagome (blue) and Tb triangular (purple) layers stacked along $c$. 
    The canted ferrimagnetic arrangement with Tb and Cr moments tilted by $\approx 25^{\circ}$ and $143^{\circ}$ with respect to the $c$ axis is adapted from neutron data~\cite{schobinger1997atomic}.  
    \textbf{b.} Actual image of the TbCr$_6$Ge$_6$ crystal.  
    \textbf{c.} Top view of the Cr kagome network, the interleaved Tb triangular sublattice, and the surrounding Ge coordination completing the HfFe$_6$Ge$_6$ structural motif.  
    \textbf{d.} Temperature-dependent susceptibility $\chi(T)$ for $\mu_0 H \parallel c$ (0.03–4 T).  
    A pronounced maximum appears near 10 K, progressively broadening and shifting to lower temperatures as the magnetic field increases.  
    \textbf{e.} For low magnetic fields ($\mu_0 H < H_c$), a sharp anomaly signals the onset of long-range magnetic order (LRO).  
    As the field approaches $H_c$, this feature broadens and shifts to lower temperatures, evolving into a crossover scale associated with short-range order (SRO).  
    At higher temperatures, the SRO regime connects smoothly to the paramagnetic (PM) phase.  
    The points shown correspond to fits of Eq.~\ref{eq:sigmoid} to the temperature-dependent susceptibility data, following the procedure described in Ref.~\cite{gokhfeld2025new}. Smooth curve is a guide to the eye.}
  \label{fig:1}
\end{figure*}

The ferrimagnetic kagome metal TbCr$_6$Ge$_6$ provides a particularly sensitive magnetic environment for transport studies. As shown in Fig.~\ref{fig:1}(a), it undergoes a ferrimagnetic transition at $T_c \approx 8$--$10\,\mathrm{K}$, where the Tb and Cr sublattices order in an antiparallel, canted configuration~\cite{schobinger1997ferrimagnetism,brabers1994magnetic,schobinger1997atomic}. Neutron diffraction confirms this structure, revealing large Tb moments and much smaller Cr moments, while magnetisation measurements show a low saturation field ($\mu_{0}H_c \approx 0.44\,\mathrm{T}$ for $H\!\parallel\!c$) and field-dependent anomalies in $M(T)$ near $T_c$, as illustrated in Fig.~\ref{fig:1}(d)~\cite{schobinger1997atomic,schobinger1997ferrimagnetism}. The small saturation field indicates that modest applied fields readily align the spins, and the persistence of anomalies above $T_c$ demonstrates that short-range magnetic correlations survive into the paramagnetic regime, as seen in Fig.~\ref{fig:1}(e).

In metallic magnets, fluctuating and disordered local moments act as a random magnetic potential for itinerant electrons, providing an additional scattering channel beyond conventional electron--electron and electron--phonon processes~\cite{Haas1968SpinDisorder,Stankiewicz2002SpinDisorder,Kudrnovsky2012SpinDisorder}. A marked reduction of spin-disorder scattering is observed upon the onset of long-range magnetic order~\cite{li2025thermal,song2017nd,kataoka2001resistivity,vanpeskitinbergen1963spin}. The balance between magnetic and non-magnetic scattering channels shifts across the phase transition. Here, TbCr$_6$Ge$_6$ provides a phase diagram with tunable spin fluctuations, enabling a direct test of how magnetic scattering reshapes both charge and heat transport. We show that scattering associated with the two magnetic sublattices remains central at low temperatures, and that the balance between magnetic and non-magnetic scattering can be tuned by external magnetic field.

Through combined measurements of longitudinal and transverse thermal and electrical conductivities across the ferrimagnetic transition, we show that the magnetoresistance of TbCr$_6$Ge$_6$ is strongly linked to its magnetization. We present the first observation of a remarkably large, field-tunable departure from the Wiedemann--Franz law around $T_c$ and at low temperatures: both longitudinal and transverse Lorenz ratios deviate strongly from the Sommerfeld value on cooling, recover partially toward $L_{xx,xy} \!\approx\! L_0$ between 5 and 7~K, and are then suppressed well below $L_0$ while the charge response remains metallic, indicating additional heat-carrying channels beyond conventional quasiparticles. The transverse thermal Hall conductivity $\kappa_{xy}$ changes sign and evolves non-monotonically with temperature, a pattern reminiscent of competing Hall-carrying bands with opposite Berry curvature. Taken together, these results establish TbCr$_6$Ge$_6$ as a distinct kagome metal in which magnetism tunes the balance between fermionic and neutral heat carriers, providing a platform to search for exotic topological effects and chiral quasiparticles. \\

\noindent \textbf{\large Results}\\
\indent
Figure~\ref{fig:1} summarizes the structural and magnetic characteristics of \texorpdfstring{TbCr$_6$Ge$_6$}{TbCr6Ge6}, 
a member of the hexagonal HfFe$_6$Ge$_6$-type (166) family crystallizing in the space group $P6/mmm$. 
Figures~\ref{fig:1}(a-c) depict the crystal and magnetic structure, consisting of alternating Cr-based kagome layers 
and Tb triangular layers stacked along the $c$ axis. 

The magnetic configuration shown in Fig.~\ref{fig:1}(a) follows the canted ferrimagnetic arrangement reported from neutron powder diffraction 
by Schobinger-Papamantellos \textit{et~al.}~\cite{schobinger1997atomic}, 
where the Tb and Cr moments are tilted by approximately $25^{\circ}$ and $143^{\circ}$ from the $c$ axis, 
with ordered moments of 8.74~$\mu_\mathrm{B}$/Tb and 0.48~$\mu_\mathrm{B}$/Cr, respectively.

Figures~\ref{fig:1}(d) and~\ref{fig:1}(e) present the temperature-dependent magnetic susceptibility $\chi(T)$ at various fields and the corresponding field–temperature phase diagram for $H \parallel c$.
At low applied magnetic fields, $\chi(T)$ shows a well-defined phase transition associated with the onset of long-range magnetic order.

With increasing field, this transition broadens and shifts to lower temperatures, indicating a gradual suppression of the low-field ordered state.

To extract the characteristic temperatures, we fitted the $\chi(T)$ curves using an asymmetric Boltzmann-type sigmoid function~\cite{gokhfeld2025new}:
\begin{equation}
\chi(T) = \chi_i +
\frac{\chi_f - \chi_i}
{\left[ 1 + \nu\,\exp\!\left(-\tfrac{T - T_\mathrm{C}}{T_w}\right) \right]^{1/\nu}},
\label{eq:sigmoid}
\end{equation}
where $\chi_i$ and $\chi_f$ are the low- and high-temperature limits, 
$T_w$ is the width parameter, and $\nu$ controls the asymmetry. The extracted temperatures are summarized in the field--temperature phase diagram in Fig.~\ref{fig:1}(e). For $\mu_0H < \mu_0H_{c}$, the critical temperature exhibits only a weak field dependence, decreasing slightly and defining the long-range-ordered (LRO) region, whereas for $\mu_{0}H > \mu_0H_{c}$, the characteristic temperature increases with field and marks the short-range-ordered (SRO) regime; at higher temperatures, the system crosses into the paramagnetic (PM) phase.

\begin{figure*}[t]
    \centering
    \includegraphics[width=\textwidth]{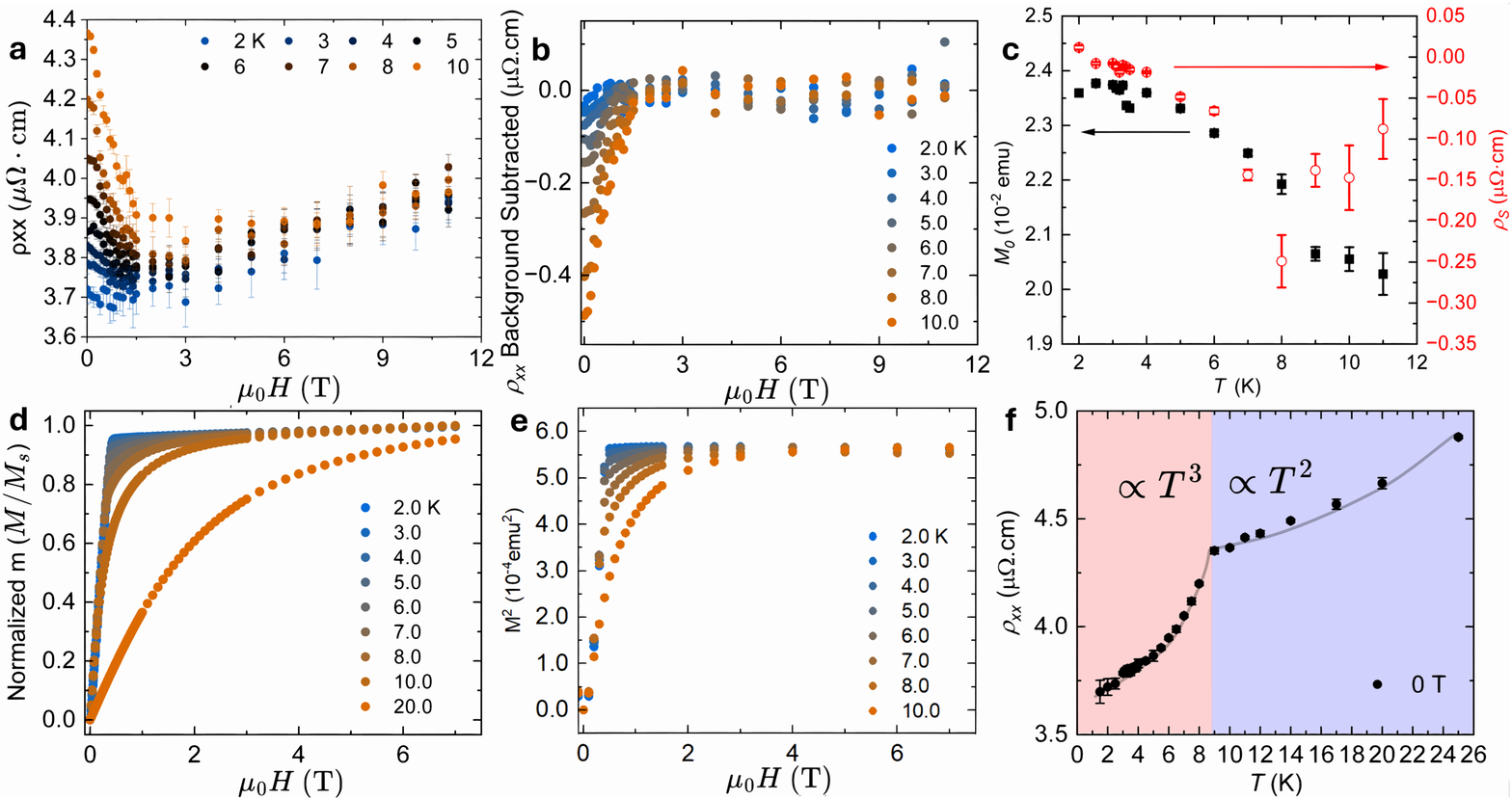}
    \caption{
    \textbf{Field-dependent electronic transport and magnetization of \TbCrGe.}
    \textbf{a.} Longitudinal electronic resistivity $\rho_{xx}(H)$ from 2--10~K showing suppression of spin-dependent scattering with increasing field.  
    \textbf{b.} Background-subtracted longitudinal resistivity $\rho_{xx}(H)$ revealing a central low-field peak consistent with spin-dependent scattering.
    \textbf{c.} Temperature dependence of the parameters $M_{0}$ (black squares) and $\rho_{S}$ (red circles) obtained from fits to $\rho_{xx}(H,M)=\rho_{xx}^{(2c)}(H)+A_M M^{2}+\rho_{S}\tanh[k(M^{2}-M_{0}^{2})]$. Here, $\rho_{xx}^{(2c)}(H)$ represent the classical two-carrier (electron--hole) background that depends on $H$ but not on $M$~\cite{Yao2025YbFe6Ge6, Zhu2018BismuthValley}. The term $A_M M^{2}$ accounts for the continuous reduction of spin-disorder scattering with increasing magnetization in $\rho_{xx}$, while the heuristic sigmoidal contribution $\rho_{S}\tanh[k(M^{2}-M_{0}^{2})]$ phenomenologically captures a crossover between scattering regimes associated with different magnetic ordered regions with distinct resistivities, where $M_{0}$ sets the characteristic magnetization scale for the crossover and $k$ controls its sharpness (see SI for more details). The progressive reduction of $M_{0}$ toward the magnetic transition near 8--10~K is consistent with the decreasing magnetization scale associated with the crossover between ferrimagnetic and paramagnetic regimes. The pronounced extremum in $\rho_{S}$ marks the temperature range in which the two-regime contribution is strongest; above the transition, $\rho_{S}$ decreases rapidly and becomes negligible. Arrows denote the corresponding axes.
    \textbf{d.} Normalized magnetization $m(H)$ displaying strong anisotropy and rapid saturation above 0.5~T.
    \textbf{e.} $M^{2}(H)$ curves.
    \textbf{f.} Temperature dependence of $\rho_{xx}$ at 0~T highlighting two regimes: below $\sim 9$~K, $\rho_{xx}\propto T^{3}$ consistent with single-magnon scattering in the ferrimagnetically ordered state, and above $\sim 9$~K, $\rho_{xx}\propto T^{2}$ reflecting coherent electronic scattering in the paramagnetic metallic regime. Smooth curves are guides to the eye. Error bars represent one standard deviation ($1\sigma$).}
    \label{fig:Fig2}
\end{figure*}

To probe the field- and temperature-dependent charge response of the magnetic state, we performed 4-probe longitudinal and Hall measurements on a single crystal of \TbCrGe\ using Lakeshore M81 lock-in system inside an Oxford Instruments Proteox MX dilution refrigerator (see Methods).

Figure~\ref{fig:Fig2} summarizes the field evolution of the electrical transport and magnetization in \TbCrGe\ between 2 and 20~K.
The longitudinal electrical resistivity $\rho_{xx}(H)$ [Fig.~\ref{fig:Fig2}(a)] shows a strong initial decrease with field at all temperatures, with the largest change occurring at low fields and a progressively weaker dependence at higher fields.

To isolate the magnetization-dependent processes in the resistivity, we subtract the high-field background that is primarily associated with the two-carrier contribution to the resistivity~\cite{Yao2025YbFe6Ge6, Zhu2018BismuthValley}.
As seen from the two-carrier expressions for $\rho^{xx}_{2c}(H)$ and $\rho^{xy}_{2c}(H)$ (see the caption Figure~\ref{fig:Fig2} and the SI for further information of the resistivity components), this background produces a smooth but strongly $H$-dependent magnetoresistance that is not linked to the evolution of the magnetization.
Removing it exposes the intrinsic low-field contributions arising from spin-dependent scattering and the crossover between magnetic ordered regions with distinct resistivities [Fig.~\ref{fig:Fig2}(b)].

The normalized magnetization $m(H)$ [Fig.~\ref{fig:Fig2}(d)] rises sharply at low fields and saturates above $\mu_{0}H \approx 0.5$~T for $H \parallel c$, consistent with a strong easy-axis response.
The corresponding $M^{2}(H)$ curves are shown in Fig.~\ref{fig:Fig2}(e), providing a quantitative measure of how the magnetization evolves with field and offering a natural reference for interpreting the field dependence of the resistivity.

A comparison between the background-subtracted $\rho_{xx}(H)$ in Fig.~\ref{fig:Fig2}(b) and the $M^{2}(H)$ curves in Fig.~\ref{fig:Fig2}(e) shows that, between 2 and 10~K, the field evolution of the low-field resistivity closely tracks the behavior of $M^{2}(H)$.
This correspondence indicates that the dominant low-field contribution to $\rho_{xx}(H)$ is strongly linked to magnetization-dependent scattering mechanism.

The temperature dependence of the parameters $M_{0}$ and $\rho_{S}$ obtained from the $\rho_{xx}(H, M)$ fits (see Fig.~\ref{fig:Fig2} caption) is shown in Fig.~\ref{fig:Fig2}(c). The crossover scale $M_{0}$, associated with the system magnetization, decreases monotonically as the system approaches the magnetic transition in the 8--10~K range, reflecting the crossover between the ferrimagnetic and paramagnetic regimes. Concurrently, the pronounced negative extremum in $\rho_{S}$, a resistivity associated with different magnetically ordered regions with distinct local resistivities (see SI) occurs near the transition temperature $T_{c}$, being progressively suppressed and becoming negligible above the transition.

The temperature dependence of the zero-field resistivity $\rho_{xx}(T)$ [Fig.~\ref{fig:Fig2}(f)] reveals two distinct transport regimes.
For $T \gtrsim 10$~K, $\rho_{xx}(T)$ follows a $T^{2}$ dependence characteristic of a paramagnetic Fermi liquid dominated by electron--electron scattering.
Below the ferrimagnetic transition near 8--10~K, $\rho_{xx}(T)$ instead follows an approximate $T^{3}$ dependence, as expected in ordered magnetic metals where scattering from low-energy magnetic excitations becomes important ~\cite{Furukawa2000JPSJ}.
The crossover between these behaviors occurs smoothly near the transition temperature.

Taken together, the conductivity, magnetization, and scaling analyses demonstrate that TbCr$_6$Ge$_6$ remains metallic throughout the 2--20~K range, with its resistivity strongly correlated with both the applied magnetic field and the evolution of the magnetization.

\begin{figure*}[t]
    \centering
    \includegraphics[width=\textwidth]{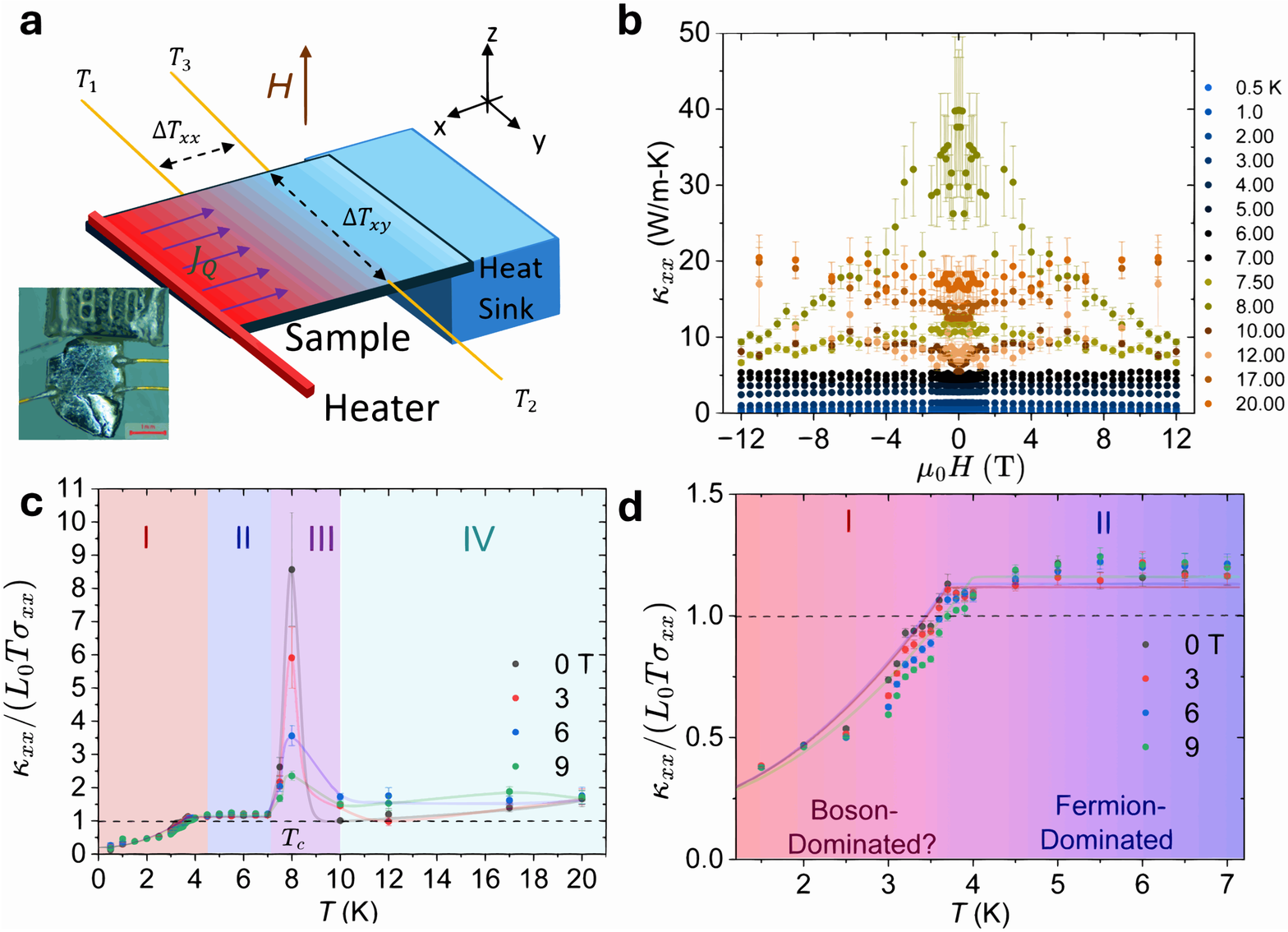}
    \caption{
    \textbf{Thermal transport and Wiedemann--Franz analysis of \TbCrGe.}
    \textbf{a.} Sketch of the steady-state thermal transport experiment. Three thermometers are connected
    to the sample via Au wires to measure longitudinal ($\Delta T_{xx}$) and transverse ($\Delta T_{xy}$) temperature gradients.
    The inset photograph shows the single-crystal device mounted with heater and thermometers inside the dilution refrigerator.
    \textbf{b.} Field dependence of the longitudinal thermal conductivity $\kappa_{xx}(H)$ at selected temperatures from 0.5 to 20~K.
    A broad enhancement centered at zero field is most pronounced near $T_c$ and is reduced with increasing field.
    \textbf{c.} Temperature dependence of the normalized longitudinal Lorenz ratio
    $L_{xx}/L_0 = \kappa_{xx}/(L_0 T \sigma_{xx})$ for $\mu_0 H = 0$, 3, 6, and 9~T.
    The shaded regions~I–IV indicate distinct regimes discussed in the text:
    I. a low-temperature regime ($T \lesssim 4$–5~K) where $L_{xx}/L_0$ is strongly reduced; 
    II. an intermediate regime ($\sim$5–7~K) where $L_{xx}/L_0$ approaches unity;
    III. a peak region around the ferrimagnetic transition temperature $T_c$ where $L_{xx}/L_0$ is strongly enhanced;
    and IV. a higher-temperature regime ($T \gtrsim 10$~K) in which $L_{xx}/L_0$ is positive and varies only weakly with temperature.
    The smooth curves are guides to the eye.
    \textbf{d.} Magnified view of normalized Lorenz ratio $L_{xx}/L_0$ in I and II regions, emphasizing the continuous evolution from the low-temperature regime
    with strongly suppressed normalized Lorenz ratio toward a higher-temperature regime that can be viewed as an extension of region~II.
    The shaded background qualitatively indicates these limiting regimes. The smooth curves are guides to the eye. Error bars represent one standard deviation ($1\sigma$).
    }
    \label{fig:Fig3}
\end{figure*}

Having established that the electrical resistivity of \TbCrGe\ is metallic and strongly correlated with the magnetization (Fig.~\ref{fig:Fig2}), we next examine the thermal transport and its comparison to the electronic channel via the Wiedemann--Franz (WF) law that allows us to search for 
thermally active carriers besides electrons that can be unconventional.

Figure~\ref{fig:Fig3} presents the temperature- and field-dependent thermal transport using a Lakeshore M81 lock-in setup mated to Stanford Research SR560 differential preamplifiers in a Proteox Dilution refrigerator under high vacuum (see Methods and SI). Fig.~\ref{fig:Fig3}(a) shows the steady-state configuration used to resolve both longitudinal and transverse temperature gradients. 

The field-dependent longitudinal thermal conductivity $\kappa_{xx}(H)$ [Fig.~\ref{fig:Fig3}(b)] exhibits a stark contrast to the electrical resistivity. The data is dominated by a pronounced low-field enhancement that is strongest near  $T_c \approx 8$–10~K. 
This enhancement appears close to the critical regime around $T_c$ identified in Fig.~\ref{fig:1}. The data is clear proof that the thermal response is sensitive to the fluctuations of the magnetic state.
$\kappa_{xx}(H)$ is suppressed at high fields as shown in Fig.~\ref{fig:Fig3}(b).
The suppression of the thermal conductivity at higher fields is consistent with moments getting increasingly field-polarized with diminishing contributions to the thermal conductivity.

The temperature evolution of the normalized longitudinal Lorenz ratio $\kappa_{xx}/(L_0 T \sigma_{xx})$ [Fig.~3(c)] shows several distinct regimes, indicated by regions I–IV.
For $T \lesssim 4$–5~K (region~I), the ratio falls well below unity, reaching $L_{xx}/L_0 \approx 0.2$ as $T\!\to\!0$ at the rate of $T^2$. At these lowest temperatures, the low Lorenz ratio indicates a substantial and surprising freezing of heat carriers both from the Cr and the Tb layers.
Within the WF framework, such low values indicate that the heat current is decoupled from measured charge  conductivity. Therefore, modified heat-carrying channels must be present. Given that no indications of this transition are present in electronic transport, here the dominant carriers are largely charge-neutral.
Between $T \approx$ 5-7~K (region~II), $L_{xx}/L_0$ is close to unity for all fields. This observation is compatible with a regime in which the WF law provides a reasonable description of the longitudinal transport and the heat channel is predominantly fermionic.
Near $T_c \approx 8$–10~K (region~III), $L_{xx}/L_0$ develops a dramatic peak as a function of temperature.
This peak coincides with the ferrimagnetic critical point, and establishes that longitudinal thermal carriers are strongly affected by the spin fluctuations.
At higher temperatures, $T \gtrsim 10$~K (region~IV), in the paramagnetic phase, the normalized  longitudinal Lorenz ratio recovers to $L_{xx}/L_0 \approx 1$ and varies only weakly with temperature over the measured field range, consistent with a paramagnetic metallic regime in which electrons and phonons account for the heat transport.

A magnified view of the low-temperature regime in I and II regions [Fig.~\ref{fig:Fig3}(d)] highlights the detailed evolution of $\kappa_{xx}(T)$ across regions ~I and ~II.
Below $\sim 4.5$~K, $\kappa_{xx}(T)$ at low fields is well described by an approximate $T^3$ dependence (Phase I).
A $T^3$ law is often used empirically to describe contributions from neutral bosonic excitations such as acoustic phonons or magnons in ordered magnets ~\cite{Zhao_Thermal_2015,Li_Thermal_2025}.
As the temperature increases above $\sim 5$~K, $L_{xx}/L_0$ increases smoothly toward unity for a range of fields (Phase II).  Between regimes I and II, the monotonic  change of $L_{xx}/L_0$ with temperature indicates a crossover, rather than an abrupt transition [Fig.~\ref{fig:Fig3}(d)], as expected from thermal excitations to a higher energy band. 

\begin{figure*}[t!]
    \centering
    \includegraphics[width=0.995\textwidth]{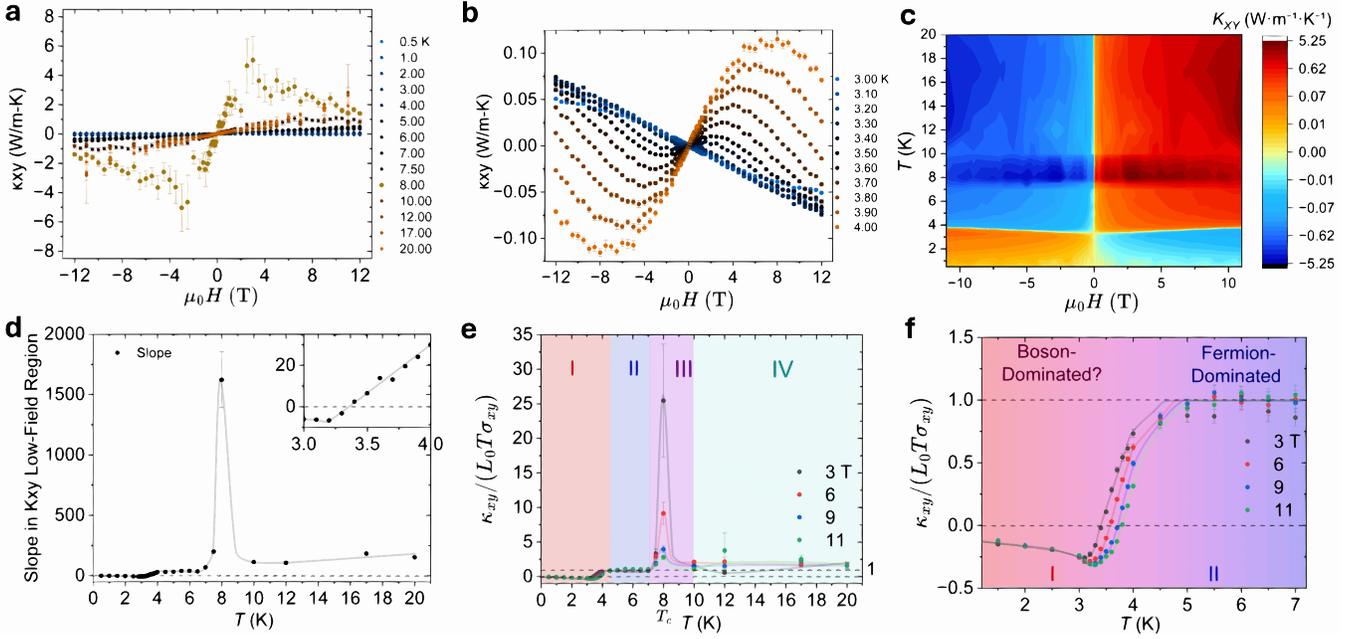}
    \caption{
    \textbf{Transverse thermal transport and normalized Lorenz analysis of \TbCrGe.}
    \textbf{(a, b)} Field-dependent thermal Hall conductivity $\kappa_{xy}(H)$ from 0.5--20~K, with a 3--4~K magnified view showing a smooth evolution from negative to positive slopes and a nonlinear regime near 8~K. 
    \textbf{(c)} Temperature--field map of $\kappa_{xy}(T,H)$ displaying a strongly antisymmetric response with sign reversals.
    \textbf{(d)} Temperature dependence of the low-field $\kappa_{xy}(H)$ slope, peaking near the ferrimagnetic transition $T_c$ (inset: low-$T$ sign change).
    \textbf{(e)} Normalized transverse Lorenz ratio $L_{xy}/L_0  = \kappa_{xy}/(L_0 T \sigma_{xy})$ for $\mu_0 H = 3$--11~T. 
    Regimes I--IV correspond to: 
    \textbf{I}, a suppressed and negative ratio ($T\lesssim4$~K); 
    \textbf{II}, a crossover toward unity (4--7~K); 
    \textbf{III}, an enhanced response near $T_c$; 
    \textbf{IV}, a weakly $T$-dependent, positive ratio above $\sim10$~K. 
    \textbf{(f)} Low-temperature enlargement of (e).
    Smooth curves are guides to the eye. Error bars represent one standard deviation ($1\sigma$).
    }
    \label{fig:Fig4}
\end{figure*}

Figure~\ref{fig:Fig4} summarizes the evolution of the transverse thermal Hall response of \TbCrGe\ across temperature and field.
At high temperatures, $\kappa_{xy}(H)$ [Fig.~\ref{fig:Fig4}(a,b)] is nearly linear with a small positive slope.
Upon cooling, $\kappa_{xy}(H)$ shows extraordinary features - the slope reverses sign between phases I and II, as shown in the expanded sweeps of Fig.~\ref{fig:Fig4}(b).
This change occurs gradually with temperature, as seen earlier for $\kappa_{xx}$ in Fig.~\ref{fig:Fig3}(d), indicating a crossover to a regime between two different thermal carriers.
The sign reversal reveals that at least two heat-carrying channels with opposite chiralities coexist at low temperatures, likely coming from two different energy bands.

The temperature--field map in Fig.~\ref{fig:Fig4}(c) presents the full evolution of the transverse thermal Hall conductivity. At higher temperatures or under strong magnetic fields, the electronic contributions re-emerge, consistent with the reduction of critical magnetic fluctuations at elevated temperatures or their quenching by large fields.
Sign-reversal around $3$–$4$~K marks the crossover between competing contributions to transverse heat transport with opposite chirality.
As the system warms, the magnitude of the $\kappa_{xy}$ peak increases, and a horizontal stripe between $8$ and $10$~K highlights a maxima in $\kappa_{xy}$ at the ferrimagnetic–paramagnetic criticality, mirroring the observation from $\kappa_{xy}$.

Further insight comes from the normalized transverse Lorenz ratio $L_{xy}/L_0$, shown in Figs.~\ref{fig:Fig4}(d,e).
Its temperature dependence separates naturally into four regimes (I–IV), consistent with $\kappa_{xx}(H)$.
In regime~I ($T \lesssim 4$~K), $L_{xy}/L_0$ is strongly suppressed and often negative, demonstrating that the dominant low-temperature transverse channel is largely decoupled from the electronic Hall conductivity and that a non-electronic contribution plays a major role.
In regime~II (4–7~K), the  normalized transverse Lorenz ratio crosses zero and approaches values of order unity, signaling the emergence of an increasingly electronic response and a reduction of the non-electronic channel.
In regime~III, near the ferrimagnetic transition ($T_c \approx 8$–10~K), $L_{xy}/L_0$ exhibits a pronounced maximum, which reflects that emergent heat carriers in this regime have the same chirality as their counterparts in regime II. Regime~IV ($T \gtrsim 10$~K) appears to be a high temperature continuation of regime II.
Fig.~\ref{fig:Fig4}(f) magnifies the low-temperature portion of this evolution, highlighting the continuous progression from negative to positive $L_{xy}/L_0$ as the balance between transverse channels shifts.\\

\noindent\textbf{\large Discussion}
\\
\indent
Comprehensive transport and thermodynamic measurements on \TbCrGe\ reveal a kagome ferrimagnet in which the normalized longitudinal and transverse Lorenz ratios are strongly temperature- and field-dependent. At high temperatures and large fields, $\rho_{xy}$ and $\kappa_{xy}$ are nearly linear in field and both $L_{xx}/L_0$ and $L_{xy}/L_0$ approach values close to unity, consistent with a regime where heat and charge are carried predominantly by quasiparticles that approximately follow Wiedemann--Franz expectations. The pronounced linear WF violation upon cooling through $T_c$ is a mystery. Also, at lower temperature, both the normalized Lorenz ratios deviate strongly from unity: $L_{xx}/L_0$ falls to $\sim 0.2$ at low fields, while $L_{xy}/L_0$ develops pronounced, sign-changing structure, all while the charge response remains metallic. These trends indicate a marked decoupling between heat and charge transport and point to unconventional charge-neutral heat-carrying channels.

The transverse response provides direct evidence for at least two contributions of opposite effective sign to the thermal Hall effect. The low-field slope of $\kappa_{xy}(H)$ is negative below $\sim 3$~K, crosses through zero near $T \approx 3.3$~K, and becomes positive at higher temperatures, with $|\kappa_{xy}|$ largest at low temperatures and intermediate fields. Second, we see a crossover between a bosonic and a fermionic behavior, with a sign changing chirality. The phenomenology we discover here is primarily magnetic. 

Experimentally, $\rho_{xx}\!\propto\!T^2$ and $L_{xx}/L_0 \approx 1$ for $T \gtrsim 10$~K, while within the ordered phase $\rho_{xx}(T)$ crosses over toward an approximate $T^3$ dependence and $L_{xx}/L_0$ is strongly suppressed as $T\!\to\!0$. A $T^2$ dependence is characteristic of a paramagnetic Fermi-liquid metal dominated by electron–electron scattering, whereas a $T^3$ law in an ordered magnet has been discussed in connection with magnon-related scattering processes~\cite{Akimoto2000}. In \TbCrGe, the simultaneous emergence of an approximate $T^3$ behavior in $\rho_{xx}$ and a large reduction of $L_{xx}/L_0$ suggests that low-energy spin excitations provide an efficient inelastic channel for heat transport and scattering. 

Finally, we consider \TbCrGe\ within the broader landscape of transverse thermal transport in kagome magnets. 
Sign-reversing thermal Hall responses have been reported in systems such as Cu(1,3-bdc), where competing magnon branches with opposite Berry curvature yield alternating contributions to $\kappa_{xy}$~\cite{Hirschberger2015}. 
Theoretical studies of kagome spin models with Dzyaloshinskii--Moriya interactions similarly predict multiple magnon modes with distinct topological character and alternating Chern flux~\cite{Laurell2018}. 
Topological magnon bands have also been observed via inelastic neutron scattering in van der Waals ferromagnets such as CrSiTe$_3$ and CrGeTe$_3$~\cite{Zhu2021TopMagnon}.

The weak rise of $\kappa_{xx}$ above 10 K indicates a relatively negligible phonon involvement. However, the sign change of $\kappa_{xy}$, the strong sensitivity of $L_{xy}/L_0$ to both field and temperature, and the pronounced low-temperature suppression of the normalized Lorenz ratio all point to the participation of multiple magnetic excitation channels in the transverse thermal response in \TbCrGe. The data suggest the presence of low-energy spin excitations, potentially carrying contrasting Berry curvatures.

A definitive identification of topological magnons in \TbCrGe\ will require momentum-resolved spectroscopy and detailed spin-wave modeling, which we expect the present study would motivate. Nonetheless, the combination of metallic charge transport, strong spin fluctuations, and field- and temperature- tunable deviations from Wiedemann--Franz scaling positions \TbCrGe\ as a promising platform for exploring magnetism-driven modifications of transverse heat and charge transport in kagome-based materials.
\\

\noindent \textbf{\large METHODS}\\
\\
\textbf{CRYSTAL GROWTH AND STRUCTURAL CHARACTERIZATION}\\
\indent
Single crystals of \TbCrGe\ were grown by a high-temperature metallic flux method. High-purity elements (Tb 99.9\%, Cr 99.95\%, Ge 99.999\%, Metal-flux 99.999\%; Alfa Aesar) were loaded into a 2~mL alumina crucible in a molar ratio of 1:1:6:20 (Tb:Cr:Ge:Metal-flux), sealed under vacuum in a fused-silica ampoule, heated to \SI{1050}{\celsius} over \SI{2}{h}, held for \SI{10}{h}, and cooled to \SI{550}{\celsius} at \SI{3.2}{\celsius\per\hour}. At \SI{550}{\celsius}, the ampoule was centrifuged to decant excess metallic flux; residual surface flux was removed by dipping in concentrated HCl. Hexagonal plate-like crystals up to $\sim$1~mg with the $c$ axis normal to the facets were obtained.

 Single-crystal X-ray diffraction (SCXRD) was performed at 293~K using Mo K$\alpha$ radiation ($\lambda = 0.71073$~\AA). The structure was solved and refined in the hexagonal space group $P6/mmm$, yielding lattice parameters $a = 5.1795(1)$~\AA{} and $c = 8.2953(4)$~\AA, consistent with the HfFe$_6$Ge$_6$-type structure.
Refinement factors were $R=0.0140$ and $wR_2=0.0311$ for 236 reflections, demonstrating high structural quality.\\

\noindent\textbf{DC MAGNETIZATION  MEASUREMENT }\\
\indent
Magnetization and susceptibility were measured using a Quantum Design  magnetic property measurement
system (MPMS-3). Temperature-dependent $\chi(T)$ was recorded under zero-field-cooled (ZFC) and field-cooled (FC) conditions for $H\!\parallel\!c$ and $H\!\parallel\!ab$, over various magnetic fields. Isothermal $M(H)$ curves were collected between \SI{2}{K} and \SI{20}{K}. \\

\noindent\textbf{ELECTRONIC TRANSPORT EXPERIMENTAL SETUP}\\
\indent
Single crystals of \TbCrGe\ with dimensions of approximately $0.5\times0.8\times0.15~\mathrm{mm^3}$ were used for electrical transport studies. 
Measurements were carried out in both an Oxford Instruments Proteox MX dilution refrigerator and a Quantum Design PPMS for consistency checks. 
For the data presented in Figs.~2–4, electrical and thermal transport were both measured inside the Proteox MX dilution refrigerator ensuring high fidelity in the determination of the normalized Lorenz ratios. 
Gold wires (\SI{2}{mil}) were bonded between Au pads and sample contacts using silver epoxy (Epo-Tek~H20E) in a standard four-probe configuration. 
Longitudinal and Hall voltages were measured simultaneously and symmetrized or antisymmetrized under $H\!\to\!-\!H$ to eliminate geometric offsets. 
Drive currents were minimized to suppress Joule heating. 

Each sample was fixed on a sapphire plate with N-grease to ensure good thermal contact and electrical isolation from the copper puck. 
Photolithographically patterned Ti/Au (\SI{10}{nm}/\SI{100}{nm}) electrodes on sapphire provided mechanically stable, low-impedance contacts with negligible thermal load. 
The electronic-transport stage was aligned beneath the thermal-transport platform to maintain a shared thermal environment for simultaneous measurement.\\

\noindent\textbf{THERMAL TRANSPORT EXPERIMENTAL SETUP}\\
\indent
Thermal transport was performed on single crystals ($0.4\times0.9\times0.1~\mathrm{mm^3}$) inside the Proteox MX dilution refrigerator using a one-heater, three-thermometer steady-state configuration, in the same high-vacuum environment with a base pressure of $\sim 0.02$~Pa. 
One end of the sample was anchored to a sapphire support with Ag paint (SPI~05001--AB) on the copper heat bath thermally linked to the mixing chamber. 
Gold wires (\SI{2}{mil}) were attached with Ag epoxy (Epo-Tek~H20E) to define longitudinal and transverse temperature leads near the sample center. 
A \SI{1}{\kilo\ohm} thin-film resistor heater and three calibrated Lakeshore RX102A thermometers (\SI{1}{\kilo\ohm}) were mounted on these leads with Ag paint. 

Low-thermal-conductivity stainless-steel wires (\SI{1}{mil}) and Manganin leads (\SI{70}{\micro\meter}) were connected to the heater and thermometers, respectively, and routed through a custom-designed PCB before being connected to copper wires extending to the feedthroughs. 
The steady heating power $P$ applied to the warm end was adjusted such that the sample temperature rise remained below 10\% of the bath temperature. 
To ensure field alignment, both thermal and electronic transport samples were mounted on a Cu plate coupled to an Attocube rotator and aligned via the front control panel of the Attocube after cooldown to \SI{4}{K}.  

Longitudinal ($\Delta T_x$) and transverse ($\Delta T_y$) temperature gradients were recorded in steady state under stepped magnetic fields. The heater current was supplied by a low-noise DC source (Keithley~6221). For each temperature and field point, the final several data points were averaged to avoid transient heating or eddy-current artifacts. Thermal and thermal Hall resistivities were obtained at fixed temperature with $H\!\parallel\!z$ and $-\nabla T\!\parallel\!x$. 
The heat bath temperature stability was better than \SI{1}{mK}, monitored using Lakeshore~350 and 372N controllers. 
The sapphire substrate electrically insulated the sample from the Cu base, ensuring that the intrinsic Cu thermal Hall contribution was negligible.\\
\\
\\
\noindent\textbf{DIFFERENTIAL METHOD AND THERMOMETER CALIBRATION}\\
\indent
Thermometer resistances $R_{T1}$, $R_{T2}$, and $R_{T3}$ were measured using the Lakeshore~M81-SSM system at $f\!\approx\!\SI{11}{Hz}$. To enhance signal-to-noise ratio and minimize environmental drift, the transverse channel was measured differentially: the resistance difference $\Delta R=R_{T2}-R_{T3}$ was amplified via a differential amplifier, SR560, ($\times100$) before detection with a lock-in amplifier from which $R_{T3}$ was calculated as $R_{T3}+\Delta R/100$. All the thermometers were calibrated \textit{in situ} as functions of temperature and field~\cite{hirschberger2017thesis}. With this method, the resolution of the transverse temperature difference reached $(1$–$1.5)\!\times\!10^{-4}$~K at $T\!=\!1$~K.

\bibliographystyle{naturemag}
\bibliography{reference}

\begin{thebibliography}{10}
\expandafter\ifx\csname url\endcsname\relax
  \def\url#1{\texttt{#1}}\fi
\expandafter\ifx\csname urlprefix\endcsname\relax\def\urlprefix{URL }\fi
\providecommand{\bibinfo}[2]{#2}
\providecommand{\eprint}[2][]{\url{#2}}

\bibitem{yin2022topological}
\bibinfo{author}{Yin, J.-X.}, \bibinfo{author}{Lian, B.} \& \bibinfo{author}{Hasan, M.~Z.}
\newblock \bibinfo{title}{Topological kagome magnets and superconductors}.
\newblock \emph{\bibinfo{journal}{Nature}} \textbf{\bibinfo{volume}{612}}, \bibinfo{pages}{647--657} (\bibinfo{year}{2022}).
\newblock \urlprefix\url{https://doi.org/10.1038/s41586-022-05516-0}.

\bibitem{li2025electron}
\bibinfo{author}{Li, Y.} \emph{et~al.}
\newblock \bibinfo{title}{Electron correlation and incipient flat bands in the kagome superconductor cscr$_3$sb$_5$}.
\newblock \emph{\bibinfo{journal}{Nat. Commun.}} \textbf{\bibinfo{volume}{16}}, \bibinfo{pages}{3229} (\bibinfo{year}{2025}).
\newblock \urlprefix\url{https://doi.org/10.1038/s41467-025-58487-x}.

\bibitem{ye2018massive}
\bibinfo{author}{Ye, L.} \emph{et~al.}
\newblock \bibinfo{title}{Massive dirac fermions in a ferromagnetic kagome metal}.
\newblock \emph{\bibinfo{journal}{Nature}} \textbf{\bibinfo{volume}{555}}, \bibinfo{pages}{638--642} (\bibinfo{year}{2018}).
\newblock \urlprefix\url{https://doi.org/10.1038/nature25987}.

\bibitem{kang2020dirac}
\bibinfo{author}{Kang, M.} \emph{et~al.}
\newblock \bibinfo{title}{Dirac fermions and flat bands in the ideal kagome metal fesn}.
\newblock \emph{\bibinfo{journal}{Nat. Mater.}} \textbf{\bibinfo{volume}{19}}, \bibinfo{pages}{163--169} (\bibinfo{year}{2020}).
\newblock \urlprefix\url{https://doi.org/10.1038/s41563-019-0531-0}.

\bibitem{liu2018giant}
\bibinfo{author}{Liu, E.} \emph{et~al.}
\newblock \bibinfo{title}{Giant anomalous hall effect in a ferromagnetic kagome-lattice semimetal}.
\newblock \emph{\bibinfo{journal}{Nat. Phys.}} \textbf{\bibinfo{volume}{14}}, \bibinfo{pages}{1125--1131} (\bibinfo{year}{2018}).
\newblock \urlprefix\url{https://doi.org/10.1038/s41567-018-0234-5}.

\bibitem{nakatsuji2015large}
\bibinfo{author}{Nakatsuji, S.}, \bibinfo{author}{Kiyohara, N.} \& \bibinfo{author}{Higo, T.}
\newblock \bibinfo{title}{Large anomalous hall effect in a non-collinear antiferromagnet at room temperature}.
\newblock \emph{\bibinfo{journal}{Nature}} \textbf{\bibinfo{volume}{527}}, \bibinfo{pages}{212--215} (\bibinfo{year}{2015}).
\newblock \urlprefix\url{https://doi.org/10.1038/nature15723}.

\bibitem{belbase2023large}
\bibinfo{author}{Belbase, B.~P.} \emph{et~al.}
\newblock \bibinfo{title}{Large anomalous hall effect in single crystals of the kagome weyl ferromagnet fe$_3$sn}.
\newblock \emph{\bibinfo{journal}{Phys. Rev. B}} \textbf{\bibinfo{volume}{108}}, \bibinfo{pages}{075164} (\bibinfo{year}{2023}).
\newblock \urlprefix\url{https://doi.org/10.1103/PhysRevB.108.075164}.

\bibitem{zhang2024large}
\bibinfo{author}{Zhang, D.} \emph{et~al.}
\newblock \bibinfo{title}{Large oscillatory thermal hall effect in kagome metals}.
\newblock \emph{\bibinfo{journal}{Nat. Commun.}} \textbf{\bibinfo{volume}{15}}, \bibinfo{pages}{6224} (\bibinfo{year}{2024}).
\newblock \urlprefix\url{https://doi.org/10.1038/s41467-024-50336-7}.

\bibitem{xu2022chargeentropy}
\bibinfo{author}{Xu, X.} \emph{et~al.}
\newblock \bibinfo{title}{Topological charge-entropy scaling in kagome chern magnet tbmn$_6$sn$_6$}.
\newblock \emph{\bibinfo{journal}{Nature Communications}} \textbf{\bibinfo{volume}{13}}, \bibinfo{pages}{1197} (\bibinfo{year}{2022}).
\newblock \urlprefix\url{https://doi.org/10.1038/s41467-022-28796-6}.

\bibitem{Krishana1995YBCO_mfp}
\bibinfo{author}{Krishana, K.}, \bibinfo{author}{Harris, J.~M.} \& \bibinfo{author}{Ong, N.~P.}
\newblock \bibinfo{title}{Quasiparticle mean free path in {YBa$_2$Cu$_3$O$_7$} measured by the thermal hall conductivity}.
\newblock \emph{\bibinfo{journal}{Physical Review Letters}} \textbf{\bibinfo{volume}{75}}, \bibinfo{pages}{3529--3532} (\bibinfo{year}{1995}).
\newblock \urlprefix\url{https://doi.org/10.1103/PhysRevLett.75.3529}.

\bibitem{Zhang2001YBCO_enhancedKxy}
\bibinfo{author}{Zhang, Y.}, \bibinfo{author}{Tao, H.}, \bibinfo{author}{Krishana, K.}, \bibinfo{author}{Behnia, K.} \& \bibinfo{author}{Ong, N.~P.}
\newblock \bibinfo{title}{Giant enhancement of the thermal hall conductivity $\kappa_{xy}$ in the superconductor {YBa$_2$Cu$_3$O$_7$}}.
\newblock \emph{\bibinfo{journal}{Physical Review Letters}} \textbf{\bibinfo{volume}{86}}, \bibinfo{pages}{890--893} (\bibinfo{year}{2001}).
\newblock \urlprefix\url{https://doi.org/10.1103/PhysRevLett.86.890}.

\bibitem{Checkelsky2012BaKFe2As2}
\bibinfo{author}{Checkelsky, J.~G.}, \bibinfo{author}{Ong, N.~P.} \emph{et~al.}
\newblock \bibinfo{title}{Thermal hall conductivity as a probe of gap structure in multiband superconductors: The case of {Ba$_{1-x}$K$_x$Fe$_2$As$_2$}}.
\newblock \emph{\bibinfo{journal}{Physical Review B}} \textbf{\bibinfo{volume}{86}}, \bibinfo{pages}{180502} (\bibinfo{year}{2012}).
\newblock \urlprefix\url{https://doi.org/10.1103/PhysRevB.86.180502}.

\bibitem{Onose2010MagnonHall}
\bibinfo{author}{Onose, Y.} \emph{et~al.}
\newblock \bibinfo{title}{Observation of the magnon hall effect}.
\newblock \emph{\bibinfo{journal}{Science}} \textbf{\bibinfo{volume}{329}}, \bibinfo{pages}{297--299} (\bibinfo{year}{2010}).
\newblock \urlprefix\url{https://doi.org/10.1126/science.1188260}.

\bibitem{Hirschberger2015FrustratedMagnet}
\bibinfo{author}{Hirschberger, M.}, \bibinfo{author}{Krizan, J.~W.}, \bibinfo{author}{Cava, R.~J.} \& \bibinfo{author}{Ong, N.~P.}
\newblock \bibinfo{title}{Large thermal hall conductivity of neutral spin excitations in a frustrated quantum magnet}.
\newblock \emph{\bibinfo{journal}{Science}} \textbf{\bibinfo{volume}{348}}, \bibinfo{pages}{106--109} (\bibinfo{year}{2015}).
\newblock \urlprefix\url{https://doi.org/10.1126/science.1257340}.

\bibitem{Kasahara2018MajoranaQuantization}
\bibinfo{author}{Kasahara, Y.} \emph{et~al.}
\newblock \bibinfo{title}{Majorana quantization and half-integer thermal quantum hall effect in a kitaev spin liquid}.
\newblock \emph{\bibinfo{journal}{Nature}} \textbf{\bibinfo{volume}{559}}, \bibinfo{pages}{227--231} (\bibinfo{year}{2018}).
\newblock \urlprefix\url{https://doi.org/10.1038/s41586-018-0274-0}.

\bibitem{Yokoi2021HalfIntegerRuCl3}
\bibinfo{author}{Yokoi, T.} \emph{et~al.}
\newblock \bibinfo{title}{Half-integer quantized anomalous thermal hall effect in the kitaev material candidate $\alpha$-{RuCl}$_3$}.
\newblock \emph{\bibinfo{journal}{Science}} \textbf{\bibinfo{volume}{373}}, \bibinfo{pages}{568--572} (\bibinfo{year}{2021}).
\newblock \urlprefix\url{https://doi.org/10.1126/science.aay5551}.

\bibitem{Bruin2022RobustHalfQuantization}
\bibinfo{author}{Bruin, J. A.~N.} \emph{et~al.}
\newblock \bibinfo{title}{Robustness of the thermal hall effect close to half-quantization in $\alpha$-{RuCl}$_3$}.
\newblock \emph{\bibinfo{journal}{Nature Physics}} \textbf{\bibinfo{volume}{18}}, \bibinfo{pages}{401--405} (\bibinfo{year}{2022}).
\newblock \urlprefix\url{https://doi.org/10.1038/s41567-021-01501-y}.

\bibitem{Czajka2021RuCl3QO}
\bibinfo{author}{Czajka, P.}, \bibinfo{author}{Gao, S.}, \bibinfo{author}{Hirschberger, M.}, \bibinfo{author}{Ji, H.} \& \bibinfo{author}{Ong, N.~P.}
\newblock \bibinfo{title}{Oscillations of the thermal conductivity in the spin-liquid state of $\alpha$-{RuCl}$_3$}.
\newblock \emph{\bibinfo{journal}{Nature Physics}} \textbf{\bibinfo{volume}{17}}, \bibinfo{pages}{915--919} (\bibinfo{year}{2021}).
\newblock \urlprefix\url{https://doi.org/10.1038/s41567-021-01243-x}.

\bibitem{Czajka2023PlanarThermalHall}
\bibinfo{author}{Czajka, P.} \emph{et~al.}
\newblock \bibinfo{title}{Planar thermal hall effect of topological bosons in the kitaev magnet $\alpha$-rucl$_3$}.
\newblock \emph{\bibinfo{journal}{Nature Materials}} \textbf{\bibinfo{volume}{22}}, \bibinfo{pages}{36--41} (\bibinfo{year}{2023}).
\newblock \urlprefix\url{https://doi.org/10.1038/s41563-022-01397-w}.

\bibitem{czajka2021oscillations}
\bibinfo{author}{Czajka, P.} \emph{et~al.}
\newblock \bibinfo{title}{Oscillations of the thermal conductivity in the spin-liquid state of $\alpha$-rucl3}.
\newblock \emph{\bibinfo{journal}{Nature Physics}} \textbf{\bibinfo{volume}{17}}, \bibinfo{pages}{915--919} (\bibinfo{year}{2021}).
\newblock \urlprefix\url{https://doi.org/10.1038/s41567-021-01243-x}.

\bibitem{Uehara2022PhononThermalHallMetallicSpinIce}
\bibinfo{author}{Uehara, T.}, \bibinfo{author}{Ohtsuki, T.}, \bibinfo{author}{Udagawa, M.}, \bibinfo{author}{Nakatsuji, S.} \& \bibinfo{author}{Machida, Y.}
\newblock \bibinfo{title}{Phonon thermal hall effect in a metallic spin ice}.
\newblock \emph{\bibinfo{journal}{Nat. Commun.}} \textbf{\bibinfo{volume}{13}}, \bibinfo{pages}{4604} (\bibinfo{year}{2022}).
\newblock \urlprefix\url{https://doi.org/10.1038/s41467-022-32375-0}.

\bibitem{Grissonnanche2020ChiralPhonons}
\bibinfo{author}{Grissonnanche, G.} \emph{et~al.}
\newblock \bibinfo{title}{Chiral phonons in the pseudogap phase of cuprates}.
\newblock \emph{\bibinfo{journal}{Nature Physics}} \textbf{\bibinfo{volume}{16}}, \bibinfo{pages}{1108--1111} (\bibinfo{year}{2020}).
\newblock \urlprefix\url{https://doi.org/10.1038/s41567-020-0965-y}.

\bibitem{Li2020SrTiO3PhononTH}
\bibinfo{author}{Li, X.}, \bibinfo{author}{Fauqu{\'e}, B.}, \bibinfo{author}{Zhu, Z.} \& \bibinfo{author}{Behnia, K.}
\newblock \bibinfo{title}{Phonon thermal hall effect in strontium titanate}.
\newblock \emph{\bibinfo{journal}{Physical Review Letters}} \textbf{\bibinfo{volume}{124}}, \bibinfo{pages}{105901} (\bibinfo{year}{2020}).
\newblock \urlprefix\url{https://doi.org/10.1103/PhysRevLett.124.105901}.

\bibitem{Katsura2010QMagnetTheory}
\bibinfo{author}{Katsura, H.}, \bibinfo{author}{Nagaosa, N.} \& \bibinfo{author}{Lee, P.~A.}
\newblock \bibinfo{title}{Theory of the thermal hall effect in quantum magnets}.
\newblock \emph{\bibinfo{journal}{Physical Review Letters}} \textbf{\bibinfo{volume}{104}}, \bibinfo{pages}{066403} (\bibinfo{year}{2010}).
\newblock \urlprefix\url{https://doi.org/10.1103/PhysRevLett.104.066403}.

\bibitem{Han2019UndopedCuprates}
\bibinfo{author}{Han, J.~H.}, \bibinfo{author}{Park, J.-H.} \& \bibinfo{author}{Lee, P.~A.}
\newblock \bibinfo{title}{Consideration of thermal hall effect in undoped cuprates}.
\newblock \emph{\bibinfo{journal}{Physical Review B}} \textbf{\bibinfo{volume}{99}}, \bibinfo{pages}{205157} (\bibinfo{year}{2019}).
\newblock \urlprefix\url{https://doi.org/10.1103/PhysRevB.99.205157}.

\bibitem{Samajdar2019SquareNeel}
\bibinfo{author}{Samajdar, R.}, \bibinfo{author}{Scheurer, M.~S.}, \bibinfo{author}{Guo, H.} \& \bibinfo{author}{Sachdev, S.}
\newblock \bibinfo{title}{Enhanced thermal hall effect in the square-lattice n{\'e}el state}.
\newblock \emph{\bibinfo{journal}{Nature Physics}} \textbf{\bibinfo{volume}{15}}, \bibinfo{pages}{1290--1294} (\bibinfo{year}{2019}).
\newblock \urlprefix\url{https://doi.org/10.1038/s41567-019-0669-3}.

\bibitem{Chen2020NearFerroelectric}
\bibinfo{author}{Chen, J.-Y.}, \bibinfo{author}{Kivelson, S.~A.} \& \bibinfo{author}{Sun, X.-Q.}
\newblock \bibinfo{title}{Enhanced thermal hall effect in nearly ferroelectric insulators}.
\newblock \emph{\bibinfo{journal}{Physical Review Letters}} \textbf{\bibinfo{volume}{124}}, \bibinfo{pages}{167601} (\bibinfo{year}{2020}).
\newblock \urlprefix\url{https://doi.org/10.1103/PhysRevLett.124.167601}.

\bibitem{Guo2020GaugeThermalHall}
\bibinfo{author}{Guo, H.}, \bibinfo{author}{Samajdar, R.}, \bibinfo{author}{Scheurer, M.~S.} \& \bibinfo{author}{Sachdev, S.}
\newblock \bibinfo{title}{Gauge theories for the thermal hall effect}.
\newblock \emph{\bibinfo{journal}{Physical Review B}} \textbf{\bibinfo{volume}{101}}, \bibinfo{pages}{195126} (\bibinfo{year}{2020}).
\newblock \urlprefix\url{https://doi.org/10.1103/PhysRevB.101.195126}.

\bibitem{Qiang2023WFkagome}
\bibinfo{author}{Qiang, X.-B.}, \bibinfo{author}{Du, Z.~Z.}, \bibinfo{author}{Lu, H.-Z.} \& \bibinfo{author}{Xie, X.~C.}
\newblock \bibinfo{title}{Topological and disorder corrections to the transverse wiedemann--franz law and mott relation in kagome magnets and dirac materials}.
\newblock \emph{\bibinfo{journal}{Phys. Rev. B}} \textbf{\bibinfo{volume}{107}}, \bibinfo{pages}{L161302} (\bibinfo{year}{2023}).
\newblock \urlprefix\url{https://doi.org/10.1103/PhysRevB.107.L161302}.

\bibitem{brabers1994magnetic}
\bibinfo{author}{Brabers, J. H. V.~J.}, \bibinfo{author}{Buschow, K. H.~J.} \& \bibinfo{author}{de~Boer, F.~R.}
\newblock \bibinfo{title}{Magnetic properties of rcr6ge6 compounds}.
\newblock \emph{\bibinfo{journal}{Journal of Alloys and Compounds}} \textbf{\bibinfo{volume}{205}}, \bibinfo{pages}{77--80} (\bibinfo{year}{1994}).
\newblock \urlprefix\url{https://doi.org/10.1016/0925-8388(94)90769-2}.

\bibitem{schobinger1997ferrimagnetism}
\bibinfo{author}{Schobinger-Papamantellos, P.}, \bibinfo{author}{Rodr{\'\i}guez-Carvajal, J.} \& \bibinfo{author}{Buschow, K. H.~J.}
\newblock \bibinfo{title}{Ferrimagnetism and disorder in the rcr6ge6 compounds (r = dy, ho, er, y): A neutron study}.
\newblock \emph{\bibinfo{journal}{Journal of Alloys and Compounds}} \textbf{\bibinfo{volume}{256}}, \bibinfo{pages}{92--96} (\bibinfo{year}{1997}).
\newblock \urlprefix\url{https://doi.org/10.1016/S0925-8388(96)03109-X}.

\bibitem{zhang2022electronic}
\bibinfo{author}{Zhang, X.} \emph{et~al.}
\newblock \bibinfo{title}{Electronic and magnetic properties of intermetallic kagome magnets $r$v$_6$sn$_6$ (r = tb--tm)}.
\newblock \emph{\bibinfo{journal}{Phys. Rev. Mater.}} \textbf{\bibinfo{volume}{6}}, \bibinfo{pages}{105001} (\bibinfo{year}{2022}).
\newblock \urlprefix\url{https://doi.org/10.1103/PhysRevMaterials.6.105001}.

\bibitem{lee2023interplay}
\bibinfo{author}{Lee, Y.} \emph{et~al.}
\newblock \bibinfo{title}{Interplay between magnetism and band topology in the kagome magnets $r$mn$_6$sn$_6$}.
\newblock \emph{\bibinfo{journal}{Phys. Rev. B}} \textbf{\bibinfo{volume}{108}}, \bibinfo{pages}{045132} (\bibinfo{year}{2023}).
\newblock \urlprefix\url{https://doi.org/10.1103/PhysRevB.108.045132}.

\bibitem{Romaka2024}
\bibinfo{author}{Romaka, V.~V.}, \bibinfo{author}{Romaka, L.}, \bibinfo{author}{Konyk, M.}, \bibinfo{author}{Tkachuk, A.} \& \bibinfo{author}{Kaczorowski, D.}
\newblock \bibinfo{title}{Structure, bonding, and properties of rcr$_6$ge$_6$ intermetallics (r = gd--lu)}.
\newblock \emph{\bibinfo{journal}{J. Solid State Chem.}} \textbf{\bibinfo{volume}{338}}, \bibinfo{pages}{124874} (\bibinfo{year}{2024}).
\newblock \urlprefix\url{https://doi.org/10.1016/j.jssc.2024.124874}.

\bibitem{yang2024crystal}
\bibinfo{author}{Yang, X.} \emph{et~al.}
\newblock \bibinfo{title}{Crystal growth, magnetic and electrical transport properties of the kagome magnet rcr6ge6 (r = gd--tm)}.
\newblock \emph{\bibinfo{journal}{Chinese Physics B}} \textbf{\bibinfo{volume}{33}}, \bibinfo{pages}{077501} (\bibinfo{year}{2024}).
\newblock \urlprefix\url{https://doi.org/10.1088/1674-1056/ad3dcf}.

\bibitem{konyk2020electrical}
\bibinfo{author}{Konyk, M.}, \bibinfo{author}{Romaka, L.}, \bibinfo{author}{Kuzhel, B.}, \bibinfo{author}{Stadnyk, Y.} \& \bibinfo{author}{Romaka, V.}
\newblock \bibinfo{title}{Electrical transport properties of rcr$_6$ge$_6$ (r = y, gd, tb, dy, lu) compounds}.
\newblock \emph{\bibinfo{journal}{Visn. Lviv Univ., Ser. Khim.}} \textbf{\bibinfo{volume}{61}}, \bibinfo{pages}{107--113} (\bibinfo{year}{2020}).
\newblock \urlprefix\url{https://doi.org/10.30970/vch.6101.107}.

\bibitem{riberolles2023orbital}
\bibinfo{author}{Riberolles, S. X.~M.} \emph{et~al.}
\newblock \bibinfo{title}{Orbital character of the spin-reorientation transition in tbmn$_6$sn$_6$}.
\newblock \emph{\bibinfo{journal}{Nat. Commun.}} \textbf{\bibinfo{volume}{14}}, \bibinfo{pages}{2658} (\bibinfo{year}{2023}).
\newblock \urlprefix\url{https://doi.org/10.1038/s41467-023-38174-5}.

\bibitem{higo2018large}
\bibinfo{author}{Higo, T.} \emph{et~al.}
\newblock \bibinfo{title}{Large thermal hall effect in a phonon-glass ba$_3$cusb$_2$o$_9$}.
\newblock \emph{\bibinfo{journal}{Nat. Photon.}} \textbf{\bibinfo{volume}{12}}, \bibinfo{pages}{73--78} (\bibinfo{year}{2018}).
\newblock \urlprefix\url{https://doi.org/10.1038/s41566-017-0086-7}.

\bibitem{onose2010observation}
\bibinfo{author}{Onose, Y.} \emph{et~al.}
\newblock \bibinfo{title}{Observation of the magnon hall effect}.
\newblock \emph{\bibinfo{journal}{Science}} \textbf{\bibinfo{volume}{329}}, \bibinfo{pages}{297--299} (\bibinfo{year}{2010}).
\newblock \urlprefix\url{https://doi.org/10.1126/science.1188260}.

\bibitem{Hirschberger2015}
\bibinfo{author}{Hirschberger, M.}, \bibinfo{author}{Chisnell, R.}, \bibinfo{author}{Lee, Y.~S.} \& \bibinfo{author}{Ong, N.~P.}
\newblock \bibinfo{title}{Thermal hall effect of spin excitations in a kagome magnet}.
\newblock \emph{\bibinfo{journal}{Phys. Rev. Lett.}} \textbf{\bibinfo{volume}{115}}, \bibinfo{pages}{106603} (\bibinfo{year}{2015}).
\newblock \urlprefix\url{https://doi.org/10.1103/PhysRevLett.115.106603}.

\bibitem{Laurell2018}
\bibinfo{author}{Laurell, P.} \& \bibinfo{author}{Fiete, G.~A.}
\newblock \bibinfo{title}{Magnon thermal hall effect in kagome antiferromagnets with dzyaloshinskii--moriya interactions}.
\newblock \emph{\bibinfo{journal}{Phys. Rev. B}} \textbf{\bibinfo{volume}{98}}, \bibinfo{pages}{094419} (\bibinfo{year}{2018}).
\newblock \urlprefix\url{https://doi.org/10.1103/PhysRevB.98.094419}.

\bibitem{Zhuo2021}
\bibinfo{author}{Zhuo, F.}, \bibinfo{author}{Wang, Q.}, \bibinfo{author}{Cheng, C.} \& \bibinfo{author}{Li, Z.}
\newblock \bibinfo{title}{Topological phase transition and thermal hall effect in breathing kagome lattice ferromagnets with dzyaloshinskii--moriya interaction}.
\newblock \emph{\bibinfo{journal}{Phys. Rev. B}} \textbf{\bibinfo{volume}{104}}, \bibinfo{pages}{144422} (\bibinfo{year}{2021}).
\newblock \urlprefix\url{https://doi.org/10.1103/PhysRevB.104.144422}.

\bibitem{ishii2013ycr6ge6}
\bibinfo{author}{Ishii, Y.}, \bibinfo{author}{Harima, H.}, \bibinfo{author}{Okamoto, Y.}, \bibinfo{author}{Yamaura, J.-i.} \& \bibinfo{author}{Hiroi, Z.}
\newblock \bibinfo{title}{Ycr6ge6 as a candidate compound for a kagome metal}.
\newblock \emph{\bibinfo{journal}{Journal of the Physical Society of Japan}} \textbf{\bibinfo{volume}{82}}, \bibinfo{pages}{023705} (\bibinfo{year}{2013}).
\newblock \urlprefix\url{https://doi.org/10.7566/JPSJ.82.023705}.

\bibitem{schobinger1997atomic}
\bibinfo{author}{Schobinger-Papamantellos, P.}, \bibinfo{author}{Rodr{\'\i}guez-Carvajal, J.} \& \bibinfo{author}{Buschow, K. H.~J.}
\newblock \bibinfo{title}{Atomic disorder and canted ferrimagnetism in the tbcr$_6$ge$_6$ compound. a neutron study}.
\newblock \emph{\bibinfo{journal}{J. Alloys Compd.}} \textbf{\bibinfo{volume}{255}}, \bibinfo{pages}{67--73} (\bibinfo{year}{1997}).
\newblock \urlprefix\url{https://doi.org/10.1016/S0925-8388(96)02872-1}.

\bibitem{Haas1968SpinDisorder}
\bibinfo{author}{Haas, C.}
\newblock \emph{\bibinfo{journal}{Phys. Rev.}} \textbf{\bibinfo{volume}{168}}, \bibinfo{pages}{531--538} (\bibinfo{year}{1968}).
\newblock \urlprefix\url{https://doi.org/10.1103/PhysRev.168.531}.

\bibitem{Stankiewicz2002SpinDisorder}
\bibinfo{author}{Stankiewicz, J.} \& \bibinfo{author}{Bartolom{\'e}, J.}
\newblock \emph{\bibinfo{journal}{Phys. Rev. Lett.}} \textbf{\bibinfo{volume}{89}}, \bibinfo{pages}{106602} (\bibinfo{year}{2002}).
\newblock \urlprefix\url{https://doi.org/10.1103/PhysRevLett.89.106602}.

\bibitem{Kudrnovsky2012SpinDisorder}
\bibinfo{author}{Kudrnovsk{\'y}, J.}, \bibinfo{author}{Drchal, V.} \& \bibinfo{author}{Turek, I.}
\newblock \emph{\bibinfo{journal}{Phys. Rev. B}} \textbf{\bibinfo{volume}{86}}, \bibinfo{pages}{144423} (\bibinfo{year}{2012}).
\newblock \urlprefix\url{https://doi.org/10.1103/PhysRevB.86.144423}.

\bibitem{li2025thermal}
\bibinfo{author}{Li, S.}, \bibinfo{author}{Guo, S.}, \bibinfo{author}{Hoke, T.} \& \bibinfo{author}{Chen, X.}
\newblock \bibinfo{title}{Thermal transport in magnetic materials: A review}.
\newblock \emph{\bibinfo{journal}{Materials Today Electronics}} \bibinfo{pages}{100156} (\bibinfo{year}{2025}).

\bibitem{song2017nd}
\bibinfo{author}{Song, J.}, \bibinfo{author}{Bi, W.}, \bibinfo{author}{Haskel, D.} \& \bibinfo{author}{Schilling, J.~S.}
\newblock \bibinfo{title}{Evidence for strong enhancement of the magnetic ordering temperature of trivalent nd metal under extreme pressure}.
\newblock \emph{\bibinfo{journal}{Physical Review B}} \textbf{\bibinfo{volume}{95}}, \bibinfo{pages}{205138} (\bibinfo{year}{2017}).
\newblock \urlprefix\url{https://doi.org/10.1103/PhysRevB.95.205138}.

\bibitem{kataoka2001resistivity}
\bibinfo{author}{Kataoka, M.}
\newblock \bibinfo{title}{Resistivity and magnetoresistance of ferromagnetic metals with localized spins}.
\newblock \emph{\bibinfo{journal}{Physical Review B}} \textbf{\bibinfo{volume}{63}}, \bibinfo{pages}{134435} (\bibinfo{year}{2001}).
\newblock \urlprefix\url{https://doi.org/10.1103/PhysRevB.63.134435}.

\bibitem{vanpeskitinbergen1963spin}
\bibinfo{author}{van Peski-Tinbergen, T.} \& \bibinfo{author}{Dekker, A.~J.}
\newblock \bibinfo{title}{Spin-dependent scattering and resistivity of magnetic metals and alloys}.
\newblock \emph{\bibinfo{journal}{Physica}} \textbf{\bibinfo{volume}{29}}, \bibinfo{pages}{917--937} (\bibinfo{year}{1963}).
\newblock \urlprefix\url{https://doi.org/10.1016/S0031-8914(63)80182-2}.

\bibitem{gokhfeld2025new}
\bibinfo{author}{Gokhfeld, D.}, \bibinfo{author}{Koblischka, M.~R.} \& \bibinfo{author}{Koblischka-Veneva, A.}
\newblock \bibinfo{title}{New model to predict thermomagnetic properties of nanostructured magnetic compounds}.
\newblock \emph{\bibinfo{journal}{Appl. Phys. A}} \textbf{\bibinfo{volume}{131}}, \bibinfo{pages}{21} (\bibinfo{year}{2025}).
\newblock \urlprefix\url{https://doi.org/10.1007/s00339-024-08131-0}.

\bibitem{Yao2025YbFe6Ge6}
\bibinfo{author}{Yao, W.} \emph{et~al.}
\newblock \bibinfo{title}{Anomalous electrical transport in the kagome magnet ybfe$_6$ge$_6$}.
\newblock \emph{\bibinfo{journal}{Phys. Rev. Lett.}} \textbf{\bibinfo{volume}{134}}, \bibinfo{pages}{186501} (\bibinfo{year}{2025}).
\newblock \urlprefix\url{https://doi.org/10.1103/PhysRevLett.134.186501}.

\bibitem{Zhu2018BismuthValley}
\bibinfo{author}{Zhu, Z.}, \bibinfo{author}{Fauqu{\'e}, B.}, \bibinfo{author}{Behnia, K.} \& \bibinfo{author}{Fuseya, Y.}
\newblock \bibinfo{title}{Magnetoresistance and valley degree of freedom in bulk bismuth}.
\newblock \emph{\bibinfo{journal}{J. Phys.: Condens. Matter}} \textbf{\bibinfo{volume}{30}}, \bibinfo{pages}{313001} (\bibinfo{year}{2018}).
\newblock \urlprefix\url{https://doi.org/10.1088/1361-648X/aaced7}.

\bibitem{Furukawa2000JPSJ}
\bibinfo{author}{Furukawa, N.}
\newblock \bibinfo{title}{Unconventional one-magnon scattering resistivity in half-metals}.
\newblock \emph{\bibinfo{journal}{J. Phys. Soc. Jpn.}} \textbf{\bibinfo{volume}{69}}, \bibinfo{pages}{1954--1957} (\bibinfo{year}{2000}).
\newblock \urlprefix\url{https://doi.org/10.1143/JPSJ.69.1954}.

\bibitem{Zhao_Thermal_2015}
\bibinfo{author}{Zhao, Z.~Y.} \emph{et~al.}
\newblock \bibinfo{title}{Thermal conductivity of ipa-cucl$_3$: Evidence for ballistic magnon transport and the limited applicability of the {Bose--Einstein} condensation model}.
\newblock \emph{\bibinfo{journal}{Phys. Rev. B}} \textbf{\bibinfo{volume}{91}}, \bibinfo{pages}{134420} (\bibinfo{year}{2015}).
\newblock \urlprefix\url{https://doi.org/10.1103/PhysRevB.91.134420}.

\bibitem{Li_Thermal_2025}
\bibinfo{author}{Li, S.}, \bibinfo{author}{Guo, S.}, \bibinfo{author}{Hoke, T.} \& \bibinfo{author}{Chen, X.}
\newblock \bibinfo{title}{Thermal transport in magnetic materials: A review}.
\newblock \emph{\bibinfo{journal}{Mater. Today Electron.}} \textbf{\bibinfo{volume}{12}}, \bibinfo{pages}{100156} (\bibinfo{year}{2025}).
\newblock \urlprefix\url{https://doi.org/10.1016/j.mtelec.2025.100156}.

\bibitem{Akimoto2000}
\bibinfo{author}{Akimoto, T.}, \bibinfo{author}{Moritomo, Y.}, \bibinfo{author}{Nakamura, A.}, \bibinfo{author}{Furukawa, N.} \& \bibinfo{author}{Machida, A.}
\newblock \bibinfo{title}{Observation of anomalous single-magnon scattering in half-metallic ferromagnets by chemical pressure control}.
\newblock \emph{\bibinfo{journal}{Phys. Rev. Lett.}} \textbf{\bibinfo{volume}{85}}, \bibinfo{pages}{3914--3917} (\bibinfo{year}{2000}).
\newblock \urlprefix\url{https://doi.org/10.1103/PhysRevLett.85.3914}.

\bibitem{Zhu2021TopMagnon}
\bibinfo{author}{Zhu, F.} \emph{et~al.}
\newblock \bibinfo{title}{Topological magnon insulators in two-dimensional van der waals ferromagnets}.
\newblock \emph{\bibinfo{journal}{Sci. Adv.}} \textbf{\bibinfo{volume}{7}}, \bibinfo{pages}{eabi7532} (\bibinfo{year}{2021}).
\newblock \urlprefix\url{https://doi.org/10.1126/sciadv.abi7532}.

\bibitem{hirschberger2017thesis}
\bibinfo{author}{Hirschberger, M.~A.}
\newblock \emph{\bibinfo{title}{Quasiparticle Excitations with Berry Curvature in Insulating Magnets and Weyl Semimetals}}.
\newblock \bibinfo{type}{Ph.d. thesis}, \bibinfo{school}{Princeton University} (\bibinfo{year}{2017}).
\newblock \urlprefix\url{https://dataspace.princeton.edu/handle/88435/dsp010c483n033}.

\end{thebibliography}

\section*{DATA AVAILABILITY}
The data that support the findings of this study are available from the corresponding author upon reasonable request.

\section*{ACKNOWLEDGEMENTS}
A.B. and J.C. acknowledge Claudio Chamon for helpful discussions. A.B., J.C., M.N., B.B., A.U., the crystal growth, bulk characterisation, electrical measurements and the work as a whole, are supported by DOE Office of Science, Basic Energy Sciences Grant DE-SC0022986. For the thermal transport part, L.L. and J.I.V.  additionally acknowledge funding from the Quantum Science Centre, a Department of Energy National Quantum Initiative Centre managed by Oak Ridge National Laboratory. A.U. acknowledges financial support from the Department of Science and Technology (DST), Government of India, through the DST-INSPIRE Faculty Fellowship (Ref. No. DST/INSPIRE/04/2019/001664)

\section*{AUTHOR CONTRIBUTIONS}
A.B. and J.C. conceived the project with inputs from M.N., L.L. and B.B.;
M.N. synthesised and did structural characterisation of the high-quality single crystalline samples with support from B.B. and A.U.;
M.N., J.C., and B.B. conducted MPMS measurements for magnetic property characterisation with support from A.U.;
J.C. set up and performed the thermal and electronic transport measurements in the dilution refrigerator and PPMS, with support from L.L.;
L.L., S.N.Z. and J.I.V. developed the two-band magnon theory. 
J.C., L.L., and M.N. produced the first draft and all authors provided inputs.

\section*{COMPETING INTERESTS}
The authors declare no competing interests.
\end{document}